# Robust diamond-like Fe-Si network in the zero-strain $Na_xFeSiO_4$ Cathode

Z. Ye[1], X. Zhao[1], S.D. Li[2], S.Q. Wu[1,3], P. Wu[4], M.C. Nguyen[1], J.H. Guo[2], J.X. Mi[5], Z.L. Gong[6], Z.Z. Zhu[3], Y. Yang[2,6], C. Z. Wang[1], K. M. Ho[1,4]

[1]Ames Laboratory - US DOE and Department of Physics and Astronomy, Iowa State University, Ames, Iowa 50011, USA.

[2]Collaborative Innovation Center of Chemistry for Energy Materials, State Key Lab of Physical Chemistry of Solid Surface and Department of Chemistry, College of Chemistry and Chemical Engineering, Xiamen University, Xiamen, 361005, China.

[3]Department of Physics, Xiamen University, Xiamen 361005, China

[4]International Center for Quantum Design of Functional Materials (ICQD), Hefei National Laboratory for Physical Sciences at the Microscale, University of Science and Technology of China, Hefei 230026, China

[5]Department of Material Science and Engineering, Xiamen University, Xiamen, 361005, China.

[6]College of Energy, Xiamen University, Xiamen, 361005, China.

**Abstract**

Sodium orthosilicates Na$_2M$SiO$_4$ ($M$ denotes transition metals) have attracted much attention due to the possibility of exchanging two electrons per formula unit. In this work, we report a group of sodium iron orthosilicates Na$_2$FeSiO$_4$, the crystal structures of which are characterized by a diamond-like Fe-Si network. The Fe-Si network is quite robust against the charge/discharge process, which explains the high structural stability observed in experiment. Using the density functional theory within the GGA+U framework and X-ray diffraction studies, the crystal structures and structural stabilities during the sodium insertion/deinsertion process are systematically investigated. The calculated average deintercalation voltages are in good agreement with the experimental result.

**1. Introduction**

Transition metal lithium orthosilicates Li$_2M$SiO$_4$ (M = Fe, Mn, Co) have attracted much attention as promising candidates for cathode materials in rechargeable Li-ion batteries owing to their large theoretical capacities.[1-7] However, due to limited supply and high cost of lithium, alternatives are being sought and evaluated, including their sodium analogues, transition metal sodium orthosilicates Na$_2M$SiO$_4$. The possibility of exchanging two electrons per formula unit (f.u.) is an outstanding property of the orthosilicates, which results in a high theoretical capacity of ~275 mA h g$^{-1}$ and ~330 mA h g$^{-1}$ for Na and Li systems, respectively. In 2011, Duncan et al. [8] first successfully synthesized Na$_2$MnSiO$_4$ with a sol-gel method and used it as the starting material to obtain a metastable polymorph (Space group, *Pn*) of Li$_2$MnSiO$_4$ via ion exchange. Chen et al. [9] also synthesized Na$_2$MnSiO$_4$ using the same method, and measured a reversible capacity of 125 mA h g$^{-1}$, which is the first evaluation of electrochemical properties for use of sodium silicate as a cathode material for Na secondary batteries. Besides, Na$_2$CoSiO$_4$ was prepared via hydrothermal method and used as positive electrode material for sodium-ion capacitors. [10]

Recently, a new material in this group, sodium iron orthosilicate Na$_2$FeSiO$_4$, was synthesized and reported to exhibit reversible electrochemical activity of 106 mA h g$^{-1}$ [11]. This material is attractive because of the natural abundance of sodium and iron. Interestingly, the major peaks of the X-ray diffraction (XRD) pattern remain at the same positions during the charge-discharge process, indicating an anomalously high structural stability which is highly desirable for improved mechanical stability in the cycle life and safety of batteries. However, a thorough understanding of the structural stability against sodium ion insertion/deinsertion cannot be achieved without identification of the crystal structure of the material. The crystal structure is also the predominant information for material design and study of the insertion/deinsertion mechanism. In this work, we discuss the polymorphic behavior of $Na_xFeSiO_4$ and attribute the high structural stability to a robust diamond-like Fe-Si network that forms the common backbone of several families of structures that occur in this material during change of sodium occupancy.

## 2. Computational Methods

The crystal structures of $Na_2FeSiO_4$ were explored using random searching with experimental input, *i.e.*, lattice parameters determined from XRD measurements [11]. The XRD pattern from the sample of solid state reaction can be indexed to a cubic cell with $a = 7.330(3)$ Å and a space group of $F\bar{4}3m$. A total number of 100 structures were generated randomly and relaxed using first-principles calculations. But we did not find any low energy structures. However, when we lifted the space group restraint to primitive cubic, we found a low energy structure with $P2_13$ (No. 198) as shown in Fig. 1.

The first-principles calculations were carried out using the projector augmented wave (PAW) method [12] within density functional theory (DFT) as implemented in the Vienna ab initio simulation package (VASP) [13,14]. The exchange and correlation energy is treated within the spin-polarized generalized gradient approximation (GGA) and parameterized by Perdew-Burke-Ernzerhof formula (PBE) [15]. Wave functions are expanded in plane waves up to a kinetic energy cut-off of 500 eV. Brillouin zone integration was performed using the Monkhorst-Pack sampling scheme [16] over k-point mesh resolution of $2\pi \times 0.03$ Å$^{-1}$. The ionic relaxations stop when the forces on all the atoms are smaller than 0.01 eV·Å$^{-1}$. The effects due to the localization of the *d* electrons of the transition metal ions in the silicates were taken into account with the GGA + U approach of Dudarev et al. [17]. Within the GGA + U approach, the on-site coulomb term *U* and the exchange term J were grouped together into a single effective interaction parameter *U-J*. In our calculations, *U-J* values were set to 4 eV for Fe.

## 3. Results and discussions
### 3.1. Crystal structures of Na$_x$FeSiO$_4$ polymorphs and zero-strain behavior with variation of Na occupancy

The structure with space group *P*2$_1$3 was plotted in Fig. 1. The SiO$_4$ and FeO$_4$ tetrahedra are connected by sharing an O-vertex and form a three dimensional diamond-like framework. Sodium atoms are located within the channel of the framework of [FeSiO$_4$] and form another diamond sub-lattice. However, the sodium sub-lattice is disrupted when sodium ions are removed from the material during the battery charge process.

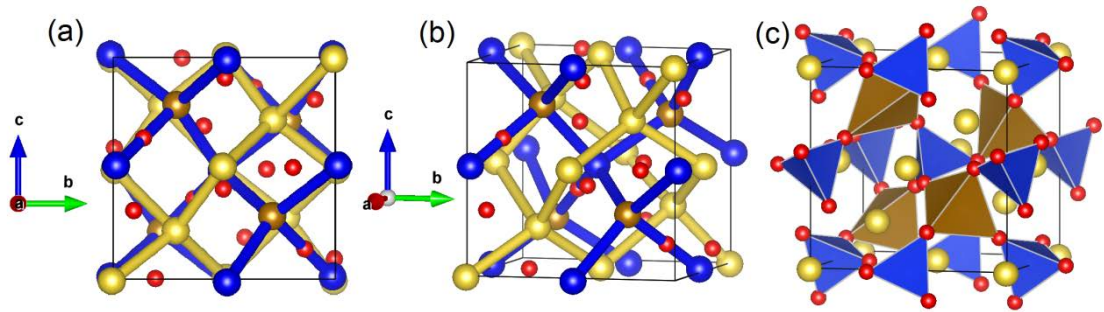

Figure 1. Crystal structure in space group *P*2$_1$3 (No. 198) that is characterized by 2 penetrating diamond sub-lattices of Na and Fe-Si cations. (a) & (b) The Na and Fe-Si networks are indicated in gold and blue, respectively. (c) Polyhedral representations of the crystal structures. The Na, Fe, Si and O atoms are shown in gold, brown , blue and red balls, respectively, whereas FeO$_4$ and SiO$_4$ tetrahedra are represent in brown and blue.

Based on the charge/discharge curves from experiments [11], the estimated number of Na atoms is 1~1.25 per f.u. in the as-produced samples, 1.5~1.75 when discharged to ~1.2 V, and 0.25~0.5 when charged to 4.5 V. To study the structural evolution of the Na$_x$FeSiO$_4$ crystal in the charge/discharge process, Na ions are removed one by one ($x = 1.75, 1.5, 1.25, ..., 0$) from the parent structures (Na$_2$FeSiO$_4$)$_4$ as shown in Fig. 1. The resulting structures are then fully relaxed using GGA+U. Figure 2(a) summarizes the cell volumes *v.s.* the energies of the lowest energy structures after removal of Na ions and relaxation from the parent structure. It is interesting to see that most of the lowest energy structures have volumes within a small scope as indicated with the grey region. Only 2 structures around the large end of the $x$ spectrum ($x \sim 2$) are located out of the grey region. This is consistent with experiments, where the estimated number of Na never reaches 0 or 2 per f.u. Figure 2(b) plots the formation energy per f.u. of Na$_x$FeSiO$_4$ *v.s.* $x$ with respect to Na$_{0.25}$FeSiO$_4$ ($x_t = 0.25$) and Na$_{1.75}$FeSiO$_4$ ($x_i = 1.75$), which is defined as [4]:

$$E_f(x) = E(Na_xFeSiO_4) - \left[\left(\frac{x-x_t}{x_i-x_t}\right)E(Na_{x_i}FeSiO_4) + \left(1 - \frac{x-x_t}{x_i-x_t}\right)E(Na_{x_t}FeSiO_4)\right] \quad (1)$$

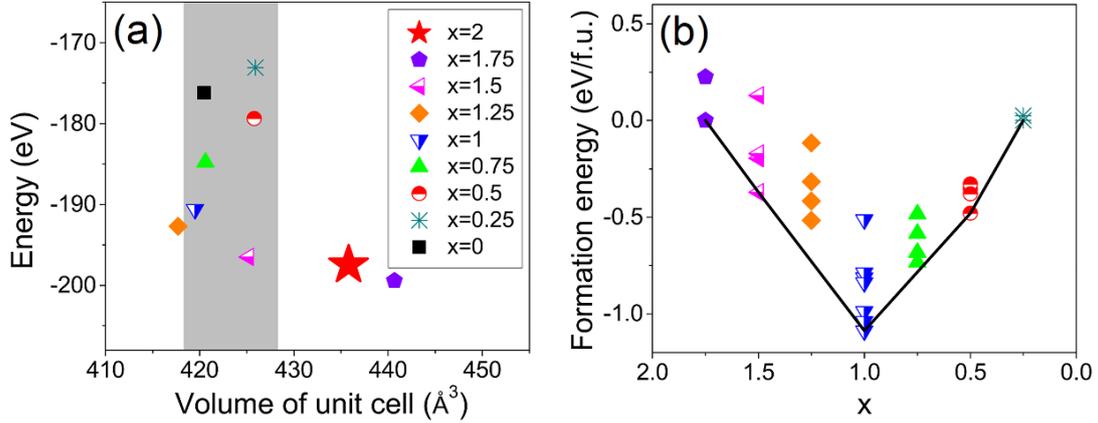

Figure 2. (a) Scatter plot of the total energy versus the volume per unit cell for the lowest energy structures with sodium ions removed from the parent structures (Na$_2$FeSiO$_4$)$_4$ (marked with red star) of space group $P2_13$. The grey region indicates the volume scope where most low energy structures locate. (b) Formation energy versus the Na composition. The solid line forms the convex hull of all the structures.

Figure 3 shows the calculated XRD patterns of the lowest energy structures for each $x$ in Fig. 2, along with the atomic structures of 4 selected configurations: $(Na_2FeSiO_4)_4$, $(Na_{1.25}FeSiO_4)_4$, $(Na_{0.75}FeSiO_4)_4$, and $(FeSiO_4)_4$ according to the $P2_13$ parent structure. It is worth noting that Fe and Si form a diamond-like backbone that is quite robust against sodium ion insertion/deinsertion. As a consequence, the material attains a high structural stability resulting in a nearly zero volume change during the charge-discharge process. This is consistent with experiment measurements [11] where the 2 major peaks at 20 and 34° (2θ) do not shift as Na ions are being extracted/re-inserted. The 2$^{nd}$ major peak at 34° (2θ) is initially broadened and eventually splits into multiple peaks. This behavior is also observed in the experimental XRD spectra [11].

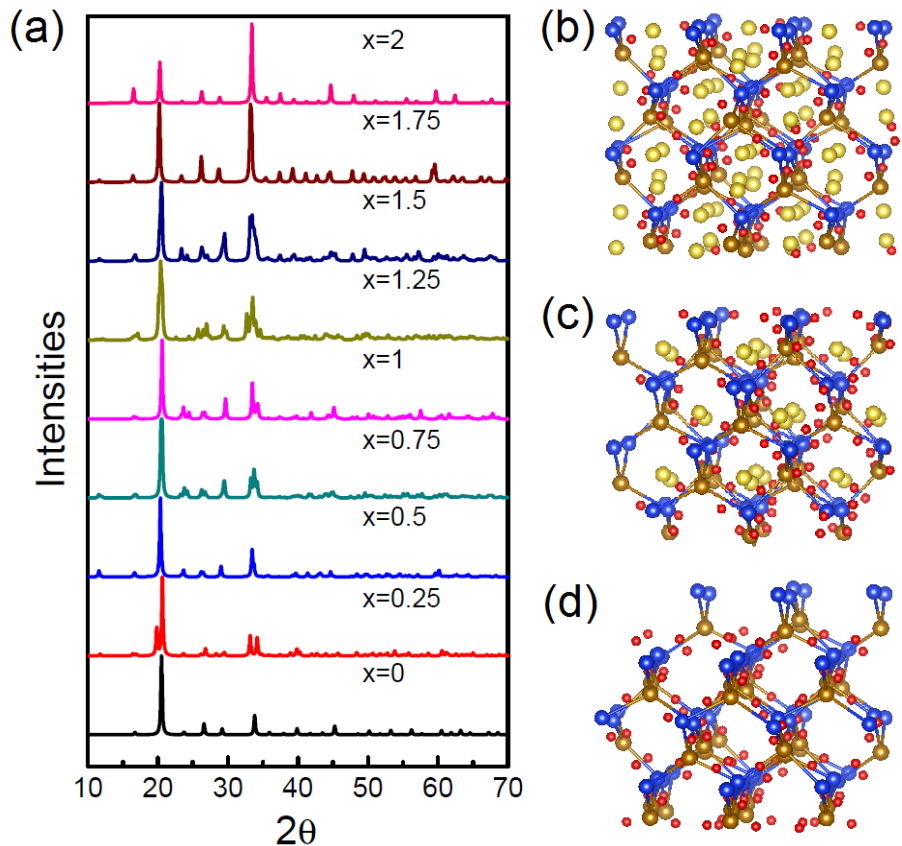

Figure 3. (a) Calculated XRD patterns of low energy $(Na_xFeSiO_4)_4$ ($x = 0$~$2$) structures obtained from extracting Na from parent $(Na_2FeSiO_4)_4$ structure of space group $P2_13$ as shown in Fig. 2(a). (b)-(d) atomic structures of $(Na_2FeSiO_4)_4$, $(NaFeSiO_4)_4$, and $(FeSiO_4)_4$, which depict a diamond-like Fe-Si network that is quite robust against sodium ion insertion/deinsertion.

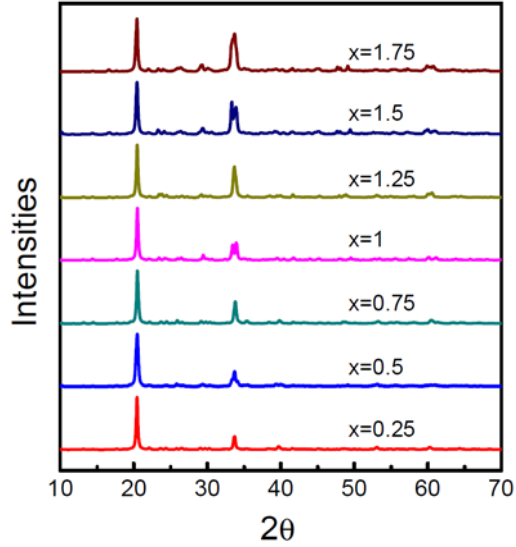

Figure 4. Calculated XRD patterns of 2×2×2 supercell structures Na$_x$FeSiO$_4$ ($x$=0.25~1.75) of space group $P2_13$.

As shown in Fig. 3(a), there are some minor peaks in the calculated XRD pattern which are not seen in the experiment. The discrepancy comes from the fact that the current modelling does not involve the disordered distribution of sodium ions when they are randomly extracted or inserted. In our modelling, we randomly remove sodium ions in ONE unit cell and relax the structure. The resulting structure inherits some characteristics from the $P2_13$ parent structure (Na$_2$FeSiO$_4$)$_4$, which results in the extra minor XRD peaks. In contrast, the real structure is an ensemble of numerous unit cells with sodium ions randomly extracted/re-inserted. The sodium ions are randomly distributed, while the Fe and Si form the aforementioned diamond-like network. The positions of oxygen anions are also disrupted to accommodate the insertion/deinsertion of sodium cations, creating disorder in the oxygen lattice. Consequently, the averaged structure of the ensemble displays a higher cubic symmetry $F\bar{4}3m$, characteristic of the robust Fe-Si network in the crystal, while the disorder in the sodium and oxygen positions remove the extra minor peaks in the experimental XRD spectra. For illustration, we approximate such random structures by constructing 2×2×2 supercells (Na$_2$FeSiO$_4$)$_{32}$ and randomly removing 0.25~1.75 sodium ions per f.u. The XRD patterns for the corresponding fully-relaxed structures are shown in Fig. 4(a). The supercell structures have similar XRD patterns with the single-cell structure, except that the extra minor peaks are greatly reduced, as expected. The minor peaks do not completely disappear, as the supercell represents a random sampling of only 8 unit cells.

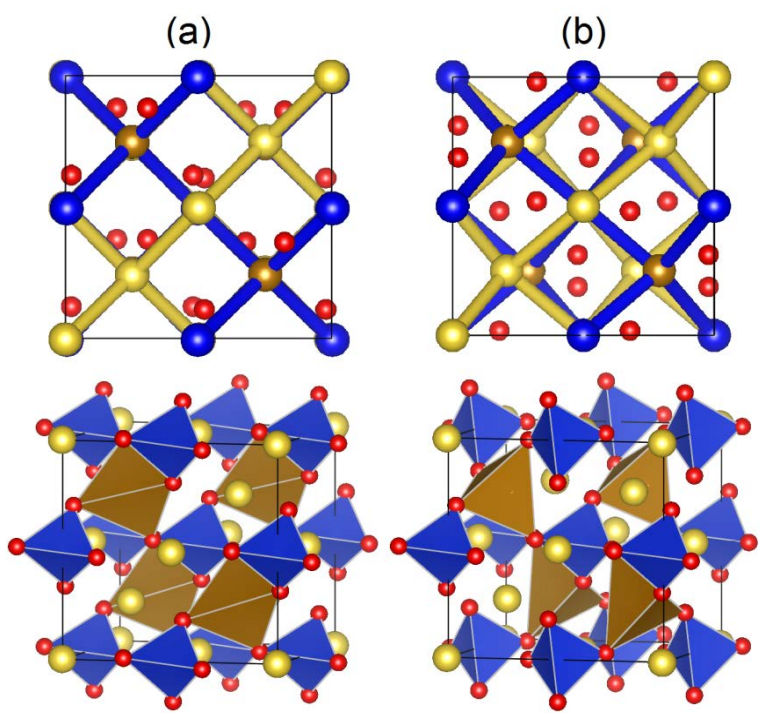

Figure 5. Crystal structures in space group (a) $C2$ (No. 5) and (b) $C222_1$ (No. 20). (top) The Na and Fe-Si networks are indicated in gold and blue, respectively. (bottom) Polyhedral representations of the crystal structures.

**3.2. Robust diamond-like Fe-Si network responsible for the zero-strain behavior**

Since the single cell structure does not necessarily have the cubic symmetry, we released the space group restraint and only enforced the cubic unit cell in the search. In this case, random searches are not effective. So we applied the motif-network scheme [18] to search possible configurations with such a cubic unit cell and no restraint of symmetries. Two low-energy structures in space group $C2$ and $C222_1$ are found after relaxation to have XRD patterns similar to experiments. They are shown in Fig. 5 (a) and (b), respectively. It can be seen from Figs. 1 and 5 that all the 3 structures ($P2_13$, $C2$, and $C222_1$) are characterized by 2 penetrating diamond sub-lattices of Na and Fe-Si cations. However, the different local arrangements of $FeO_4$ and $SiO_4$ tetrahedrons result in slightly different orientations of the diamond sub-lattices in the 3 different structures.

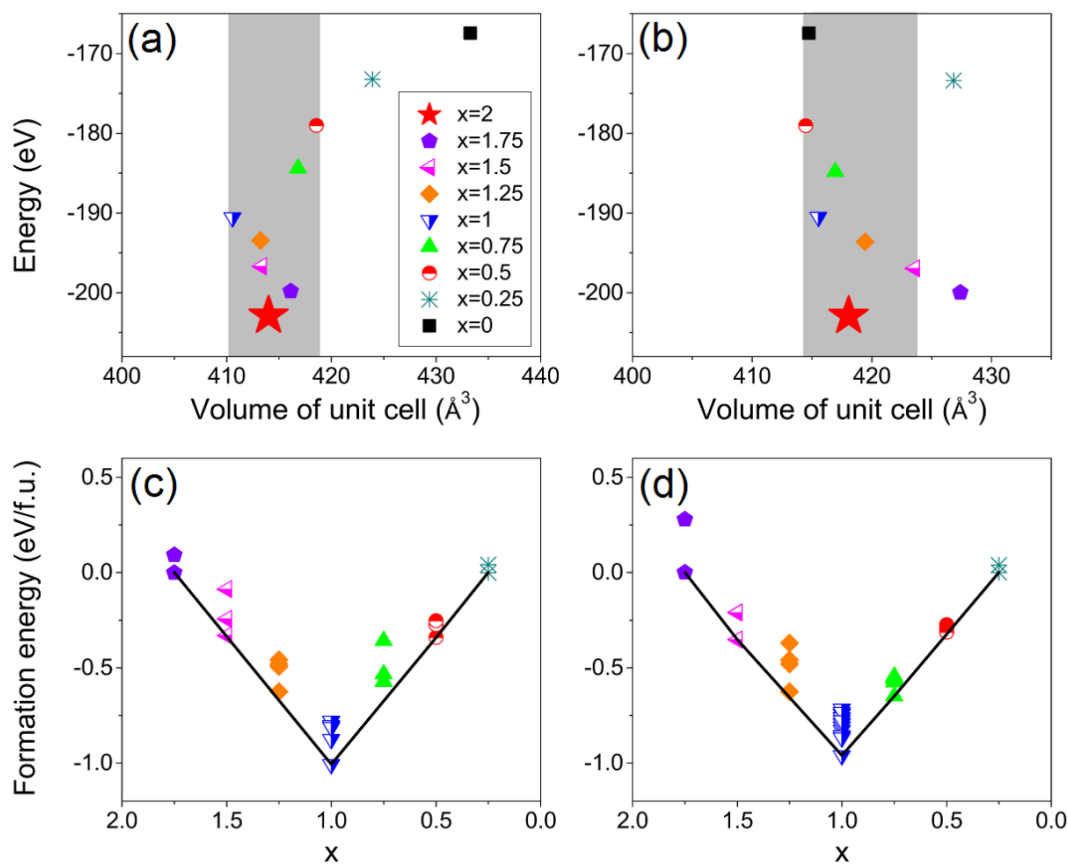

Figure 6. (a) & (b) Scatter plot of the total energy versus the volume per unit cell for the lowest energy structures with sodium ions removed from the parent structures $(Na_2FeSiO_4)_4$ (marked with red star) of space group $C2$ and $C222_1$, respectively. The grey region indicates the volume scope where most low energy structures locate. (c) & (d) Formation energy versus the Na composition for $C2$ and $C222_1$ structures, respectively. The solid line forms the convex hull of all the structures.

We repeat the calculation as shown in Fig. 2 for the 2 new structures and plot the results in Fig. 6. It can be seen from Fig. 2 and Fig. 6 that the parent $P2_13$ structure has a much higher energy and a larger cell volume than the $C2$ and $C222_1$ structures to accommodate the $FeO_4$ and $SiO_4$ tetrahedrons that satisfy the higher symmetry.

However, the lowest energy structures after removal of Na ions has almost the same energies and cell volumes for all the 3 structures, indicating a common Fe-Si network underlying the 3 different parent structures through the Na insertion/deinsertion. This robust Fe-Si diamond framework contributes to the nearly zero volume change during the charge-discharge process.

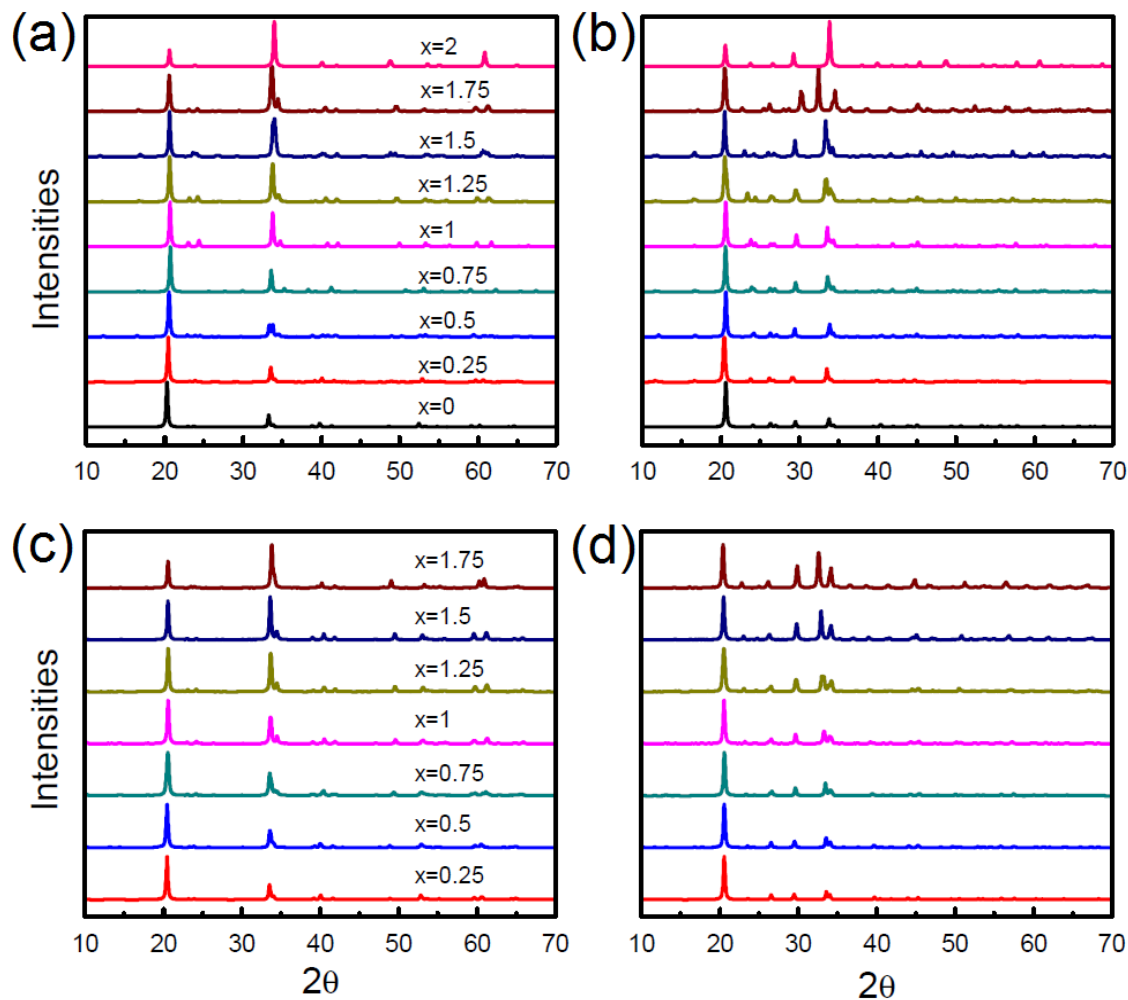

Figure 7. (a) & (b) Calculated XRD patterns of low energy $(Na_xFeSiO_4)_4$ ($x$ = 0~2) structures of space group $C2$ and $C222_1$, respectively. (c),(d) Calculated XRD patterns of 2×2×2 supercell structures $Na_xFeSiO_4$ ($x$ = 0.25~1.75) of space group $C2$ and $C222_1$, respectively.

Figure 7(a) and (b) shows the calculated XRD patterns of the lowest energy structures in Fig. 6(a) and (b) with Na ions being removed from the parent $C2$ and $C222_1$ structure, respectively. The positions of the 2 major peaks are exactly like the $P2_13$ structure. However, the $C2$ structure has fewer minor peaks than the other 2 structures. We also repeat the XRD calculation for the 2×2×2 supercell of $C2$ and $C222_1$ structure and plot the results in Fig. 7(c) and (d). In the XRD pattern of the $C2$ structure, the 2$^{nd}$ major peak is almost as sharp as the 1$^{st}$ one even in the random supercell structure. We found that Na extraction causes less disruption in the positions of the remaining Na atoms in this structure. We note that the $Na_xFeSiO_4$ sample prepared by solid state reaction also has a similar sharp 2$^{nd}$ major peak [11], while the sample synthesized by sol-gel method has a broadened 2nd major peak. Different processing methods probably create slightly different structures associated with different XRD patterns. However, these structures share a very similar Fe-Si network, as indicated from the fixed positions of the 2 major XRD peaks. As Fig. 1(a) and the top of Fig. 5(a)(b) show, the 2 penetrating diamond sub-lattices of Na and Fe-Si cations are the best neatly lined up for the $C2$ structure. This may explain why the Na extraction causes less disruption in the positions of the remaining Na atoms in this structure during discharge process. The oxygen anions are differently distributed in the 3 structures. Nevertheless, the detailed arrangement of oxygen ions cannot be determined unless XRD patterns with very high resolutions are acquired.

All the 3 $Na_2FeSiO_4$ polymorphs are similar in structure: the $SiO_4$ and $FeO_4$ tetrahedra are connected by sharing an O-vertex and Na's are located within the channel of the framework of $[FeSiO_4]$. To characterize the structural evolution in the charge/discharge process, we calculate the bond lengths of Fe-O and Si-O and distortions of $FeO_4$ and $SiO_4$ tetrahedra for the 3 $Na_xFeSiO_4$ polymorphs as functions of $x$ ($0.25 \leq x \leq 1.75$). For all 3 structures, the bond lengths increase as $x$ increases, as expected. Besides, the 3 polymorphs have similar values of distortion parameters within a reasonably small range. The detailed results can be found in the Supplementary Information.

As emphasized before, the Fe-Si diamond-like framework is responsible for the nearly zero volume change during the charge-discharge process. It is thus interesting to analyze the stability of the Fe-Si network. We calculate the bond length of Fe-Si for the 3 $Na_xFeSiO_4$ polymorphs as a function of $x$ and plot the results in the Supplementary Information. The Fe-Si bond length varies within a very small range of ~1%, which results in a small volume change during the charge-discharge process.

### 3.3. Influence of crystal structure in the electrochemical properties

The average voltages of sodium deintercalation/intercalation can be predicted with first-principles calculations. Following the well-established methods, the average deintercalation voltage V $v.s.$ Na/$Na^+$ can be calculated using the following equation,

$$V = -\frac{E_{coh}[Na_{x_2}Host] - E_{coh}[Na_{x_1}Host] - (x_2-x_1)E_{coh}[Na]}{x_2-x_1} \quad (2)$$

where $x_2$ and $x_1$ are the Na composition before and after the sodium extraction from the host, respectively. Fig. 8 shows the calculated average voltages for the 3 $Na_2FeSiO_4$ polymorphs. They have similar calculated voltages, except that the $P2_13$ structure has an additional stable structure at $x = 0.5$. As a comparison, a value of 1.9V and 4.0V are experimentally observed during the charge/discharge process, corresponding to the $Fe^{2+}/Fe^{3+}$ and $Fe^{3+}/Fe^{4+}$ redox reaction, respectively.

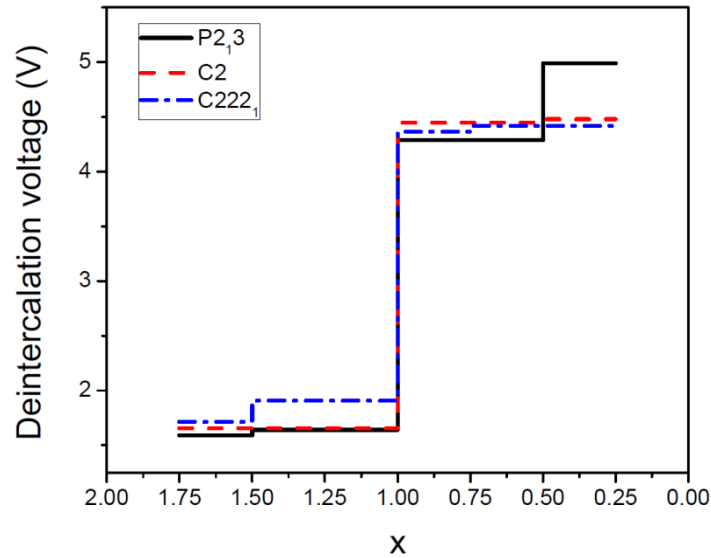

Figure 8. Deintercalation voltage versus the Na composition for $P2_13$, $C2$ and $C222_1$ structures, respectively.

**Conclusions**

In this work, we have studied the structural characteristics, the stability, and the electrochemical properties of the 3 Na$_x$FeSiO$_4$ polymorphs crystallizing in the space group *P*2$_1$3, *C*2 and *C*222$_1$, respectively. All the polymorphs are characterized by a common Fe-Si diamond-like network that is quite robust against the Na insertion/extraction, which is responsible for the nearly zero volume change observed experimentally during the charge/discharge process. Different processing methods probably create polymorphs with very similar arrangement of cations but different distribution of oxygen anions. The experimentally observed polymorphs are very stable in structure resulting in a very small volume change during the charge/discharge process, which is highly desirable for improved mechanical stability for rechargeable Na-ion batteries. The idea of the Fe-Si diamond-like network that is robust in the charge/discharge process may be generalized to other transition metal sodium orthosilicates Na$_x$MSiO$_4$ to achieve the zero-strain for cathode materials.

**Acknowledgement**

Work at Ames Laboratory was supported by the US Department of Energy, Basic Energy Sciences, Division of Materials Science and Engineering, under Contract No. DE-AC02-07CH11358, including a grant of computer time at the National Energy Research Scientific Computing Center (NERSC) in Berkeley, CA. Work at Xiamen University was supported by the National Basic Research Program of China (973 program, Grant No. 2011CB935903), the National Natural Science Foundation of China (Grants No. 21233004, 21473148 and 21021002, and in part 21428303), the Natural Science Foundation of Fujian Province of China (Grant No. 2015J01030), and the Fundamental Research Funds for the Central Universities (Grant No. 20720150034). Work at University of Science and Technology of China (USTC) was supported by the China Postdoctoral Science Foundation (Grant No. BH2340000063), Program of Introducing Talents of Discipline to Universities of Ministry of Education (MOE) & the State Administration of Foreign Experts Affairs of the People's Republic of China (SAFEA) and the Supercomputing Center of USTC.